\documentclass[twocolumn,showpacs,preprintnumbers,amsmath,amssymb]{revtex4}
\usepackage{graphicx}
\usepackage{dcolumn}
\usepackage{bm}

\graphicspath{{fig/},{fig2/}}

\begin{document}

\title{Chaotic instantons and enhancement of tunneling in double-well system with polychromatic perturbation}

\author{V.I. Kuvshinov, A.V. Kuzmin and V.A. Piatrou}
\affiliation{Joint Institute for Power and Nuclear Research - Sosny of the National Academy of Sciences,\\ Krasina str. 99, Minsk,  220109, Belarus }
\date{\today}
\begin{abstract}
Chaotic instanton approach allows to describe analytically the influence of the polychromatic perturbation on quantum properties of nonlinear systems. Double well system with single, multiple and polychromatic kicked perturbation is considered in the paper to compare quantitative analytical predictions with the results of numerical calculations. Chaotic instantons are responsible for the enhancement of tunneling far away from the exact (avoided) level crossings in framework of the approach used. Time-independent averaged system is used for regular approximation of the chaotic instanton solution in order to take into account its contribution to the ground quasienergy doublet splitting. Formula for the ground quasienergy levels' splitting was derived averaging trajectory action in the stochastic layer in the framework of chaotic instanton approach. Results of quasienergy spectrum numerical calculations and simulations for tunneling dynamics are in good agreement with the obtained analytical predictions.
\end{abstract}



\maketitle

\section{Introduction}
\addcontentsline{toc}{section}{Introduction}

Investigation of the influence of small perturbation on the behavior of the nonlinear dynamical systems continues to attract permanent interest for several last decades~\cite{Liberman, Zaslavski,Haake,Reichl}. The connection between the semiclassical properties of perturbed nonlinear systems and purely quantum processes such as tunneling is a reach rapidly developing field of research nowadays \cite{Haake,Grifoni:98}. Our insight in some novel phenomena in this field was extended during the last decades. The most intriguing among them are the chaos assisted tunneling (CAT) and the closely related coherent destruction of tunneling (CDT).

In particular, the former is an enhancement of tunneling in the perturbed low-dimensional systems at small external field strengths and driving frequencies~\cite{Lin:90,Peres:91,Plata:92,Holthaus:92}. This phenomenon takes place when levels of the regular doublet undergo an avoided crossing with the chaotic state~\cite{Bohigas:93,Latka:94}. At the semiclassical level of description one considers tunneling between KAM-tori embedded into the "chaotic sea". The region of chaotic motion affects tunneling rate because, compared to direct tunneling between tori, it is easier for the system to penetrate primarily into the chaotic region, to travel then along some classically allowed path and finally to tunnel onto another KAM-torus~\cite{Utermann:94, Mouchet:01}. The latter, CDT phenomenon, is a suppression of tunneling which occurs due to the exact crossing of two  states with different symmetries from the tunneling doublet \cite{Grossmann:91}. In this case the tunneling time diverges which means the total localization of quantum state on the initial torus.

CAT phenomenon as well as CDT were experimentally observed in a number of real physical systems. The CAT observation  between whispering gallery-type modes of microwave cavity having the form of the annular billiard was reported in  Ref.~\cite{Dembowski:00}. The same phenomenon for ultracold atoms was experimentally investigated in Refs.~\cite{Steck:01,Hensinger:01}. The study of the dielectric microcavities provided evidences for CAT as well~\cite{Podolskiy:05}. Both CAT and CDT phenomena were observed in two coupled optical waveguides \cite{vorobeichik:03, Valle:07}. Recently experimental evidence of coherent control of single particle tunneling in strongly driven double well potential was reported in Ref. \cite{kierig:08}. 

The most common methods which are used to investigate the interplay between semiclassical properties of perturbed nonlinear systems and quantum processes are numerical methods based on Floquet theory~\cite{Utermann:94,Shirley:65,Grifoni:98} and Random Matrix Theory~\cite{Leyvraz:96}. Among other approaches we would like to mention the scattering approach for billiard systems~\cite{Doron:95,Frischat:98} and approach based upon the presence of a conspicuous tree structure hidden in a complicated set of tunneling branches~\cite{Shudo:96,Shudo:98,Shudo:01}.

In this paper we will consider the original analytical approach based on instanton technique. Enhancement of tunneling in system with external force in framework of this approach occurs due chaotic instantons which appear in perturbed case. This approach was proposed in Refs. \cite{KKS:02,KKS:03,KK:05} and used in Ref.~\cite{Igarashi:06}. Chaotic instanton approach will be developed further here using averaged time-independent Hamiltonian and exploited for description of the enhancement of tunneling in the polychromatically kicked double well system. Previously polychromatic perturbation was investigated numerically only~\cite{Igarashi:06}. The main purpose of the present study is to prove the ability of developed chaotic instanton approach to give quantitative analytical description of tunneling in polychromatically perturbed systems well agreed with independent numerical calculations based on Floquet theory. It will give additional support and pulse for the further development of analytical methods to investigate tunneling phenomenon in quantum systems with mixed classical dynamics. Alternative approach based on quantum instantons which are defined using an introduced notion of quantum action was suggested in Refs.~\cite{Jirari:01, Paradis:05}. Another instanton approach was developed recently for description of ordinary, dynamical and resonant tunneling in various nonperturbed systems~\cite{Deunff:10}. Analytical approach to describe tunneling in perturbed systems based on nonlinear resonances consideration was developed in Refs.~\cite{Brodier:02,Brodier:01}. A theory for dynamical tunneling process using fictitious integrable system was recently given in Refs. \cite{Backer:08,Backer:10}.

Double well potential is a model which is convenient to use for tunneling analysis. This system is well studied in the nonperturbed case, e.g. on the base of instanton technique~\cite{Polyakov:77,Vainshtein:82} or WKB method~\cite{Zinn-Justin:81}. Double well potential is often used for description of processes which occurred in wide range of real physical systems: such as  flipping of the ammonia molecule \cite{Merzbacher}, transfer of protons along hydrogen bonds in benzoic-acid dimers at low temperatures \cite{Skinner:88,Oppenlander:89} and macroscopic quantum coherence phenomena in superconducting quantum interference devices \cite{Rouse:95,Friedman:96,Friedman:00} and nanomagnets \cite{Awschalom:92,Barco:99}. Perturbation in this paper is regarded in the form of  kicks. One of the attractive features of this type of perturbation is the extensively-investigated simple quantum map which stroboscobically evolves the system from kick $n$ to kick $n+1$. Kicked systems are recently used for experimental realization of a such novel concept as a quantum ratchet~\cite{Sadgrove:07,Dana:08}. Double kicked perturbation was investigated experimentally in Ref. \cite{Jones:04}.

The paper is divided into several sections. Chaotic instantons are analyzed using averaged time-independent Hamiltonian of the kicked system in section~\ref{sec:sys}. Results obtained by means of the averaged Hamiltonian are used in section \ref{sec:formula} to derive analytical formula for lowest  quasienergy doublet splitting dependence on perturbation parameter in single kicked system. Numerical calculations are performed to check the validity of this formula in the section~\ref{sec:num}. Multiple and polychromatic kick perturbations are considered in the sections \ref{sec:2kick} and \ref{sec:aperiod}, respectively.

\section{Chaotic instantons in kicked double-well potential}\label{sec:sys}

Hamiltonian of the particle in the double-well potential can be written down in the following form:
\begin{equation}\label{eq:H}
H_0 = \frac{p^2}{2 m} + a_0\, x^4 - a_2\, x^2,
\end{equation}
where $m$ - mass of the particle, $a_0, a_2$ - parameters of the
potential. We consider the perturbation of the kick-type and choose it as follows:
\begin{equation}\label{eq:V}
V_{per} = \epsilon\, T \, x^2 \sum^{+ \infty}_{n = - \infty} \delta(t- n T),
\end{equation}
where $\epsilon$ and $T$ are perturbation strength and period, respectively, $t$ - time. Dependence of the perturbation on coordinate was chosen in the form of $x^2$ in order to preserve spatial symmetry in the perturbed system. Hamiltonian of this system is
\begin{equation}\label{SystemHamiltonian}
H = H_0 + V_{per}.
\end{equation}

Now we implement Wick rotation ($t \rightarrow - i
\tau$) and define Euclidean Hamiltonian
\begin{equation}\label{H_E}
\mathcal{H}^E = \mathcal{H}^E_0 - \epsilon\, T \, x^2 \sum^{+ \infty}_{n = - \infty} \delta(\tau- n T),
\end{equation} 
where $\mathcal{H}^E_0$ - nonperturbed Euclidean Hamiltonian which is given by
\begin{equation}\label{H0_E}
	\mathcal{H}^E_0 = \frac{p^2}{2 m} - a_0\, x^4 + a_2\, x^2.
\end{equation}

\begin{figure}[!ht]
\centering
\includegraphics[angle = 270,width=0.48\textwidth]{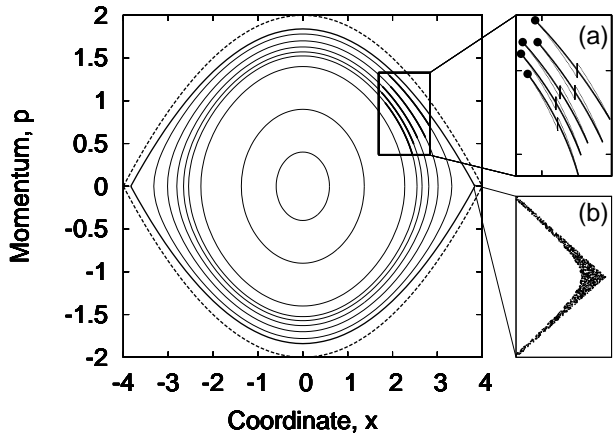}
\caption{Phase space of the system with averaged Hamiltonian with parameter $\tilde{a}_2 (\epsilon = 0.02)$. Separatrix in this system (thick solid line) and in the nonperturbed system (dashed line) are shown in the figure. Comparison of the particle classical motion on one period of the perturbation in averaged  (thick solid lines) and kicked (thin solid lines) systems from the set of initial conditions (thick points) are shown in the inset $(a)$. Inset $(b)$ shows a stroboscopic plot of chaotic trajectory of the kicked particle near the turning point of the separatrix in averaged system.}\label{fig:effps}
\end{figure}

Euclidean equations of motion of the particle in the nonperturbed double-well potential ($\epsilon = 0$) have a well known solution - instanton. This solution is used for calculation of the ground energy splitting in the system without perturbation \cite{Polyakov:77,Vainshtein:82} and explains the rate of the tunneling process in it. Another solutions of the Euclidean equations of motion besides ordinary instanton are required to explain dynamical tunneling in perturbed system. Perturbation destroys the separatrix and some trajectories in its vicinity go to infinity. Narrow stochastic layer is formed nearby the nonperturbed separatrix due to the perturbation. ``Chaotic instanton'' appears in this layer. Chaotic instanton is the closest to the destroyed nonperturbed separatrix trapped trajectory (see figure \ref{fig:effps}). Thus it plays a dominant role in tunneling in perturbed system.

Now let us construct the averaged time-independent Hamiltonian for the double well system with the perturbation of the kick-type. It is calculated using Euclidean Hamiltonian (\ref{H_E})
\begin{equation}\label{def:hameff_E}
\mathcal{H}^E_{av} = \frac{1}{T} \int^T_0 \mathcal{H}^E dt = \frac{p^2}{2 m} - a_0 x^4 + \left(a_2 - \epsilon \right)x^2.
\end{equation}

Hamiltonian (\ref{def:hameff_E}) is coincided with nonperturbed Hamiltonian (\ref{H0_E}) when parameter $a_2$ is replaced by $\tilde{a}_2$. The latter is defined as follows:
\begin{equation}\label{a2eff}
	\tilde{a}_2 (\epsilon) = a_2 - \epsilon.
\end{equation}
In contrast to the kicked system (\ref{H_E}) averaged Hamiltonian (\ref{def:hameff_E}) is time-independent. Since the form of the potential is changing, there should be restriction for the perturbation strength variation. This restriction  follows from our assumption that ordinary instanton approach should be valid for the averaged potential. Condition for ordinary instanton approach applicability~\cite{Vainshtein:82} can be written down for instanton action in averaged potential
\begin{equation}\label{seff_cond}
S_{eff} = \frac{2 \tilde{a}^{3/2}_2}{3 a_0} \gtrsim 6.
\end{equation}
Thus, for restriction considered we obtain
\begin{equation}\label{condition}
	\epsilon  \lesssim \epsilon_{max} = a_2 - 3 \sqrt[3]{3} \, a^{2/3}_0.
\end{equation} 

Figure \ref{fig:effps} shows a possibility to describe properties of the classical motion in the kicked double well system in Euclidean time using averaged system (\ref{def:hameff_E}). This Hamiltonian was used to analyze the perturbed system phase space in Euclidean time and to construct an approximation for chaotic instanton solution. This approximation is a separatrix in the averaged model. Using it's properties we obtain the formula for stochastic layer width
\begin{equation}\label{eq:dH}
	\Delta \mathcal{H}^E_s \approx \frac{\epsilon a_2}{2 a_0}.
\end{equation} 
Expression (\ref{eq:dH}) will be used in the following section in order to obtain analytical formula for the lowest  quasienergy doublet splitting dependence on the perturbation strength in the kicked system.

\section{Ground doublet quasienergy splitting formula}\label{sec:formula}

The lowest doublet energy splitting in two loop approximation in the nonperturbed double well potential is the following (see \cite{Wohler:94} and review \cite{Vainshtein:82}):
\begin{equation}\label{dE_0}
 \Delta E_0 = 2 \, \omega_0 \sqrt{\frac{6}{\pi}} \, \sqrt{S_{inst}} \, exp\left(- S_{inst} - \frac{71}{72}
\frac{1}{S_{inst}}\right),
\end{equation}
where $\omega_0$ - oscillation frequency near the bottom of the wells, $S_{inst} = 2 \sqrt{m} \, a^{3/2}_2 /(3 \, a_0)$ - nonperturbed instanton action.

Ground doublet quasienergy splitting ($\Delta \eta$) in the kicked system in the framework of our approach is expressed in terms of chaotic instanton action ($S_{ch}$) through the formula which is similarly to (\ref{dE_0}):
\begin{equation}\label{eq:dn}
	\Delta \eta = 2 \, \omega_0 \sqrt{\frac{6}{\pi}} \, \sqrt{S_{ch}} \, exp\left(- S_{ch} - \frac{71}{72}
\frac{1}{S_{ch}}\right),
\end{equation}
where chaotic instanton action can be calculated by averaging the nonperturbed trajectory action~($S(E)$) over energy for stochastic layer width
\begin{equation*}
S_{ch} = \frac{1}{\Delta \mathcal{H}^E_s}
\,\int^{E_{max}}_{E_{min}}  S(E)  d\,E = \frac{1}{\Delta \mathcal{H}^E_s} \,\int^{\Delta \mathcal{H}^E_s}_{0}  S(\xi)  d\,\xi,
\end{equation*}
where we have made a transformation to the integral over the energy difference $\xi=E_{inst}-E$ in last expression.
Using nonperturbed trajectory action expansion near the separatrix $S(E) = \pi J(E_{inst} - \xi) \approx  S_{inst} - \alpha \, \sqrt{\frac{m}{a_2}} \; \xi$  expression for chaotic instanton action can be calculated directly. Here $\alpha = (1 + 18 \ln 2)/6$ is a numerical coefficient. Thus, for chaotic instanton action we obtain
\begin{equation}\label{eq:Sch}
S_{ch} =  S_0 - \frac{\alpha}{2} \sqrt{\frac{m}{a_2}}\; \Delta
\mathcal{H}^E_s.
\end{equation}

Now we can write down analytical formula for the ground quasienergy levels splitting using expressions~(\ref{eq:dH}), (\ref{dE_0}), (\ref{eq:dn}) and (\ref{eq:Sch}):
\begin{equation}\label{an-fomula}
\Delta \eta(\epsilon) = \Delta E_0 \, e^{k \, \epsilon},
\end{equation}
where 
\begin{equation}\label{k}
	k = \frac{\alpha \, \sqrt{m \, a_2}}{4 \, a_0}.
\end{equation}

Tunneling period in the kicked double well potential is expressed in terms of ground quasienergy levels splitting
\begin{equation}\label{eq:T}
	T_{tun} = \frac{2\, \pi}{\Delta\, \eta}.
\end{equation}
Increasing of the perturbation parameter gives exponential rise to ground quasienergy splitting and to the tunneling frequency ($\nu_{tun} (\epsilon) = \Delta\, \eta (\epsilon)$). The last exponential factor in the expression~(\ref{an-fomula}) is responsible for the tunneling enhancement in the perturbed system. In nonperturbed case formula (\ref{an-fomula}) coincides with the expression~(\ref{dE_0}). Formulas~(\ref{an-fomula}) and~(\ref{eq:T}) will be checked in numerical calculations in the next section.

\section{Numerical calculations}\label{sec:num}

For the computational purposes it is convenient to choose the eigenvectors of harmonic oscillator as the basis vectors. In this representation matrices of the Hamiltonian (\ref{eq:H}) and the perturbation~(\ref{eq:V}) are real and symmetric. They have the following forms ($n \ge m$):
\begin{align*}
H^0_{m\, n} &= \delta_{m \;n} \left[\hbar \omega \left(n + \frac12\right) + \frac g 2 \left(\frac32 \, g\, a_0 \, (2 m^2 + 2  m + 1)\right. \right.\\ & -\left.\left. a'_2 (2 m + 1) \right)  \right] \\  
&+ \delta_{m + 2 \; n} \;\frac{g}{2} \left(g\, a_0  (2 m + 3) - a'_2 \right) \sqrt{(m + 1)(m  + 2)}\\
&+ \delta_{m + 4 \; n} \frac{a_0 g^2}{4} \sqrt{(m + 1)(m + 2)(m + 3)(m + 4)},\\
V_{m\, n} &= \epsilon \, T \; \frac{g}{2} \; \left(\delta_{m + 2 \; n}\; \sqrt{(m + 1)(m  + 2)} +  \delta_{m \;n} (2 m + 1)\right),
\end{align*}
where $g  = \hbar/m \omega$ and $a'_2 = a_2 + m \,\omega^2/2$, $\hbar$ is Planck constant which we put equal to $1$, $\omega$ - frequency of the basis harmonic oscillator which is arbitrary, and so may be adjusted to optimize the computation. We use the value $\omega = 0.2$ with parameters $m~=~1,$ $a_0~=~1/128,$ $a_2 = 1/4$ in most of calculations which are chosen in such a way that nonperturbed instanton action is large enough for energy splitting formula for nonperturbed system to be valid and not too big in order to decrease errors of numerical calculations. The matrix size is chosen to be equal to $200 \times 200$. Calculations with larger matrices give the same results. System of computer algebra Mathematica was used for numerical calculations.

\begin{figure}[!h]
\centering
\includegraphics[angle = 270,width=0.48\textwidth]{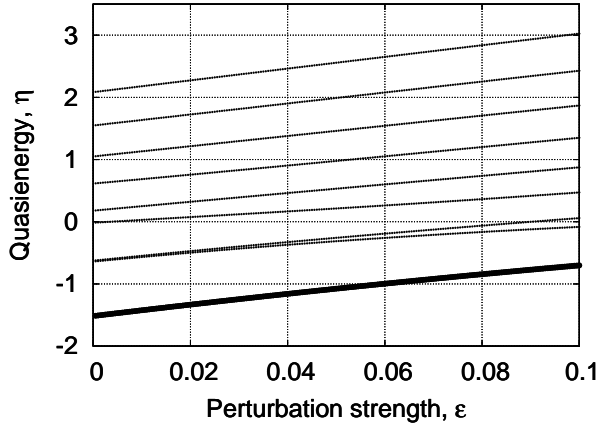}
\caption{Quasienergy spectrum for the ten lowest average energy levels. Thick lines - doublet with the minimal average energy.}\label{fig:qeS}
\end{figure}

We calculate eigenvalues of the one-period evolution operator ($e^{- i \hat{H}_0 T/2} e^{- i \epsilon T \hat{x}^2} e^{- i \hat{H}_0 T/2}$) and obtain quasienergy levels ($\eta_k$) which are related with the evolution operator eigenvalues ($\lambda_k$) through the expression $\eta_k = i \, \ln \lambda_k/T$. Then we get ten levels with the lowest one-period average energy. We obtain these levels using the formula $\left<v_i\right|H_{av}\left|v_i\right>$. Here $H_{av}$ is the averaged Hamiltonian in Minkowski space, $\left|v_i\right>$ are the eigenvectors of the one-period evolution operator. The dependence of ten lowest levels' quasienergies on the strength of the perturbation for the model parameters mentioned above is shown in the figure~\ref{fig:qeS}. Quasienergies of two levels with the minimal average energy are shown by thick lines. They are too close to each other to be resolved in the figure due to very small splitting.

\begin{figure}[!ht]
\centering
\includegraphics[angle = 270,width=0.48\textwidth]{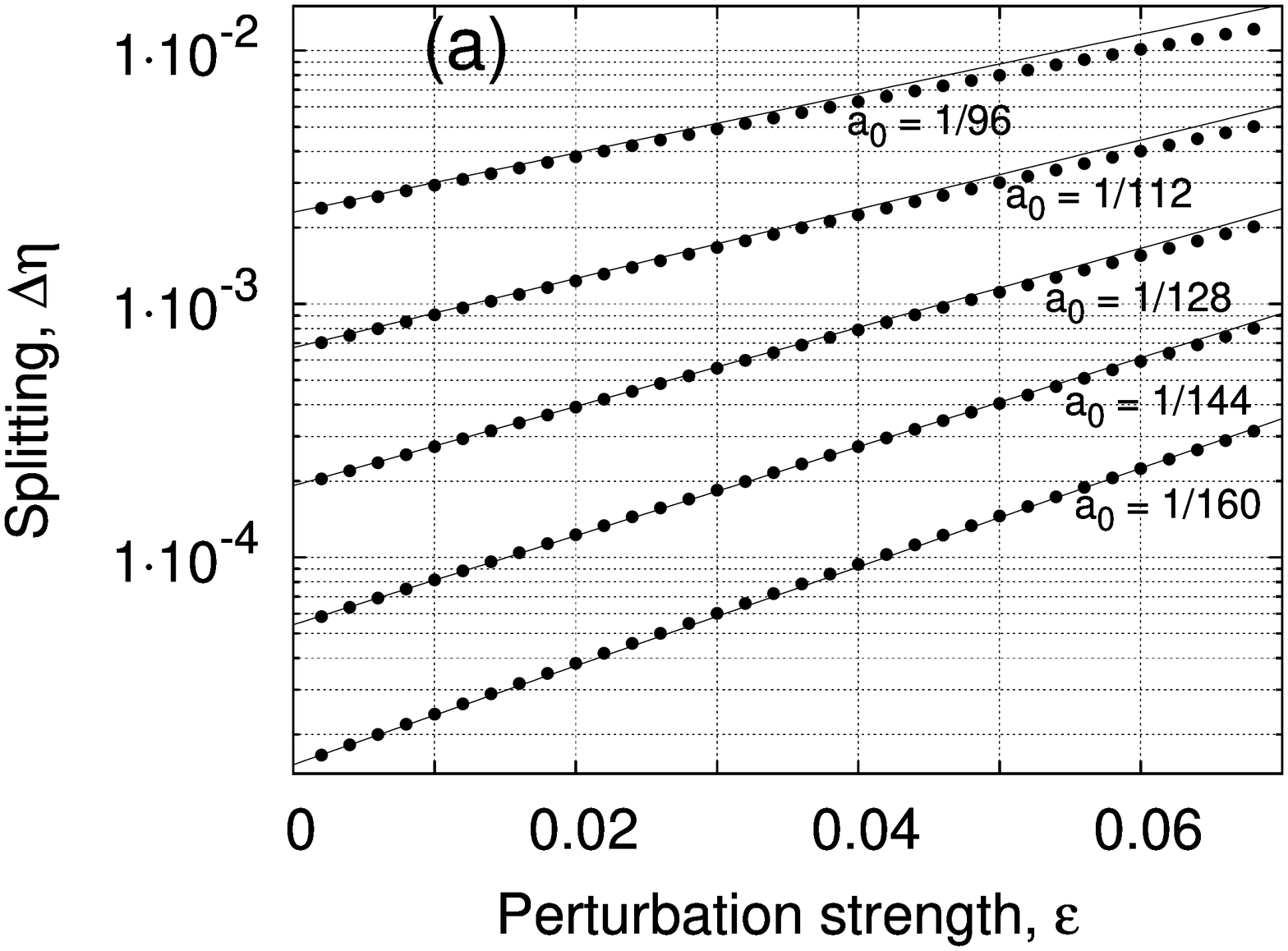}
\includegraphics[angle = 270,width=0.48\textwidth]{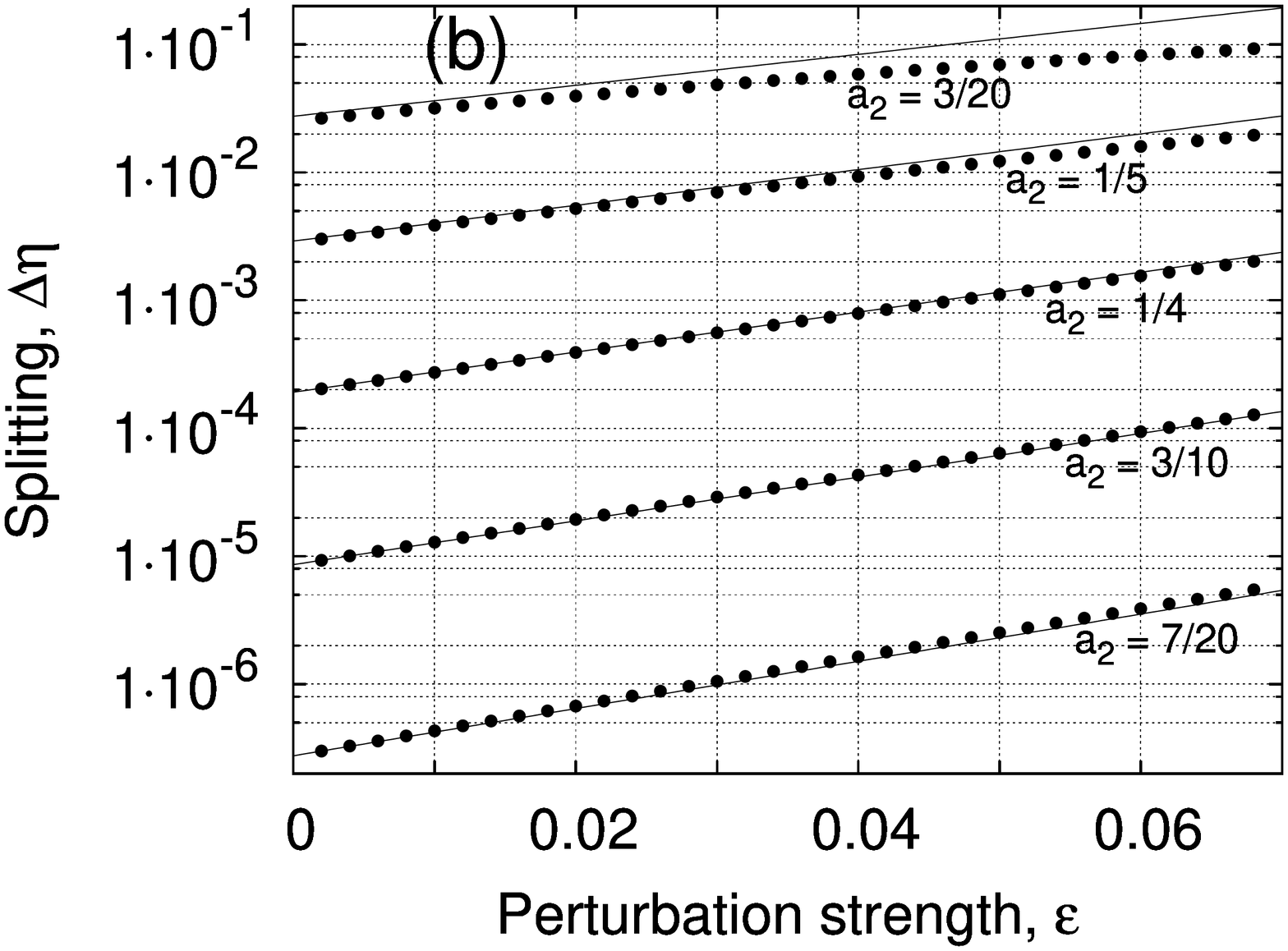}
\caption{Quasienergy splitting as a function of the strength of the perturbation for different values of model parameters $a_0$ (a) and $a_2$ (b). Lines - analytical
formula (\ref{an-fomula}), points - numerical results.}\label{fig:dE}
\end{figure}

Performed numerical calculations give the dependence of the ground quasienergy splitting on the strength of the perturbation for different values of model parameters $a_0$ (fig.\ref{fig:dE}(a)) and $a_2$ (fig.\ref{fig:dE}(b)). We fix parameter $a_2 = 1/4$ for figure \ref{fig:dE}(a) and $a_0~=~1/128$ for figure \ref{fig:dE}(b). Results of numerical calculations are plotted in the figure~\ref{fig:dE} by points. Axis $\Delta\eta$ is shown in logarithmic scale. Obtained dependencies are exponential as it was predicted by chaotic instanton approach and obtained analytical formula (\ref{an-fomula}).

Analytical results are plotted in the figures~\ref{fig:dE}~(a)~and~(b) by straight solid lines. Numerical points lie close to these lines. The agreement between numerical calculations and analytical expression is good in the parametric region considered.

\begin{figure}[!t]
\centering
\includegraphics[angle = 270,width=0.48\textwidth]{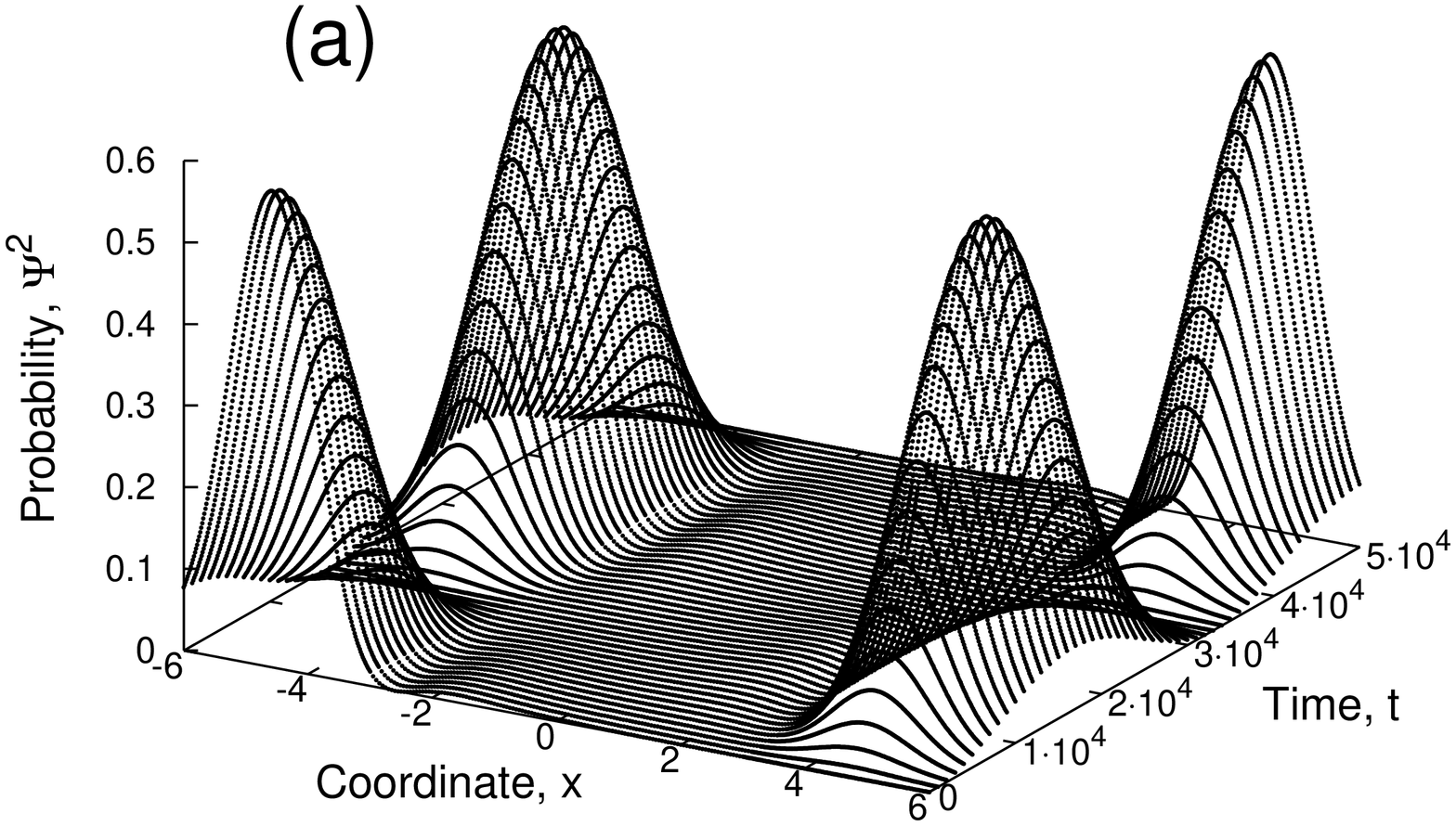}\hfil
\includegraphics[angle = 270,width=0.48\textwidth]{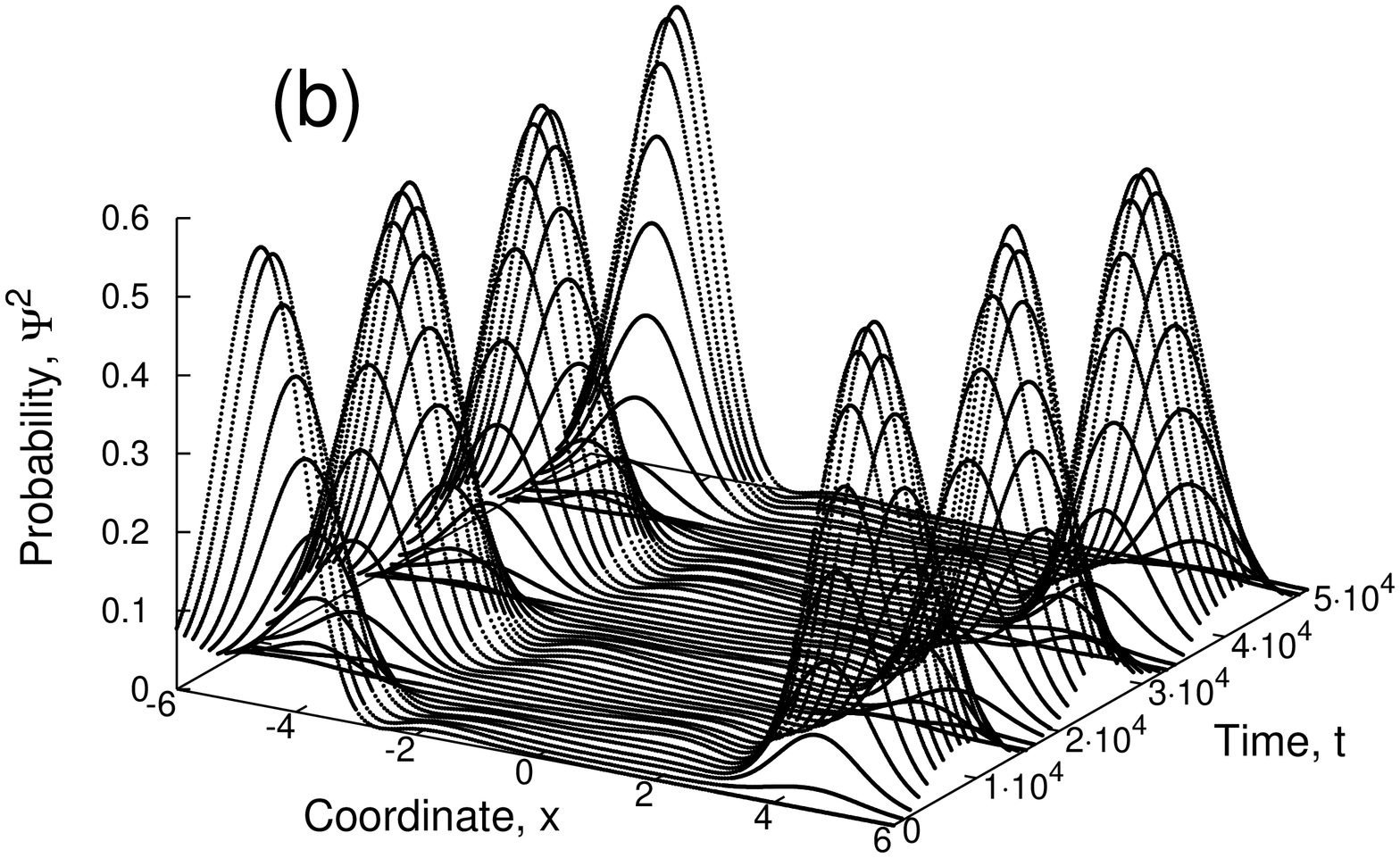}\\
\caption{Quantum mechanical tunneling in kicked double well potential. Perturbation parameters: (a) $\epsilon = 0$, (b) $T = {2 \pi}/{4}, \epsilon = 1.9 \cdot 10^{-2}$.}\label{fig:tun}
\end{figure}

Now lets perform numerical simulations for the tunneling process in the kicked double well system and check an applicability of the formulas (\ref{an-fomula}) and (\ref{eq:T}) for this process. For this purpose we regard the double well potential (\ref{eq:H}) with parameters $m~=~1,$ $a_0~=~1/128,$ $a_2 = 1/4$ and the same basis vectors as for previous calculations. We take a symmetric superposition of two lowest nonperturbed states as a initial wave packet. These packet is localized in left well of potential. Numerical simulations we provide by multiplying initial wave function by one period evolution operator. The results of numerical simulations for the two values of the perturbation strength are shown in the figure~\ref{fig:tun}. The dependence of the localization probability of the wave packet on the coordinate and time is presented in figures. Minima of the nonperturbed double well potential~(\ref{eq:H}) are situated in points $x = -4$ and $x = 4$. Tunneling between these points in nonperturbed system is demonstrated in the figure~\ref{fig:tun}(a). Evolution of the initial wave packet in perturbed case is shown in the figure~\ref{fig:tun}(b). Perturbation parameters for these simulations are $T = {2 \pi}/{4}$ and $\epsilon = 1.9 \cdot 10^{-2}$. They are chosen in such a way to speed up a tunneling in two times in comparison with nonperturbed system. Figures \ref{fig:tun}(a) and \ref{fig:tun}(b) demonstrate this enhancement. Fourier analysis of the dependence of the localization probability of the wave packet in left well on time in perturbed case confirms analytical assumptions mentioned above.

\begin{figure}[!h]
\centering
\includegraphics[angle = 270,width=0.48\textwidth]{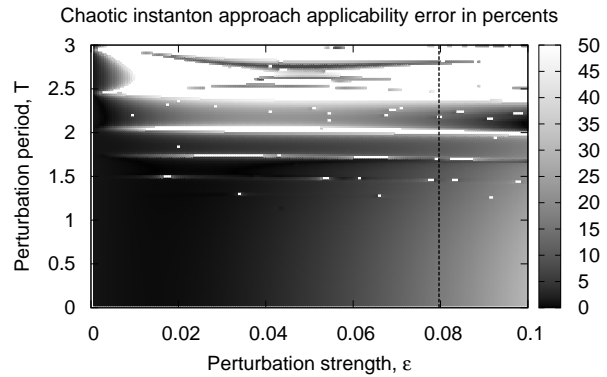}
\caption{Comparison error of the analytical formula~(\ref{an-fomula}) with results of numerical calculations in percents.}\label{fig:appl}
\end{figure}

In order to check applicability of the developed approach we carry out a series of the numerical calculations for wide range of the perturbation parameters. Results of the analysis is performed in figure~\ref{fig:appl}. Region of quantitative agreement between analytical and numerical results is shown by black color in the figure. There are two restrictions of the developed approach applicability. The first one is that
model parameters should be far away from the exact (avoided) level crossings. Thus we have the restriction for the perturbation period ($T \ll 2 \pi/{\omega_0}$, where $\omega_0$ - oscillation frequency near the bottom of the wells).
Another restriction for analytical predictions is a condition for ordinary instanton approach applicability which imply the maximum for  perturbation strength~(\ref{condition}). This maximum is shown in the figure  by dashed line. The two restrictions mentioned above explain accurately the figure~\ref{fig:appl}.

Inverse sign in the expression of the perturbation (\ref{eq:V}) will induce exponential suppression of tunneling in the system. It can be demonstrated numerically as well.

\section{Multiple kick perturbation}\label{sec:2kick}

Lets consider double-well system with multiple kick perturbation.  Hamiltonian of this system is the following:
\begin{align}
H & = \frac{p^2}{2 m} + a_0 x^4 - a_2 x^2 + \epsilon_1 T x^2 \sum^{+ \infty}_{n = - \infty} \delta(t- n T) \nonumber \\ & + \epsilon_2 T x^2 \sum^{+ \infty}_{n = - \infty} \delta(t + \Delta  T- n T),\label{ham_multi}
\end{align}
where $\epsilon_1$ and $\epsilon_2$ are strength of two perturbations, $T$ is the period for both kicking sequences, $\Delta  T$ - shift between these sequences.

Averaged time-independent Hamiltonian for the system under investigation is given by
\begin{equation}\label{def:hameff_multi}
H_{av}  = \frac{1}{T} \int^T_0 H dt = \frac{p^2}{2 m} + a_0 x^4 - \left(a_2 - \epsilon_1 - \epsilon_2 \right) x^2.
\end{equation}

Using the last expression we can rewrite restriction for perturbation strength (\ref{seff_cond}) in multiple kick case
\begin{equation}\label{condition_multi}
\epsilon_1 + \epsilon_2  \lesssim \epsilon_{max} = a_2 - 3 \sqrt[3]{3} \, a^{2/3}_0.
\end{equation}

\begin{figure}[!h]
\centering
\includegraphics[angle = 270,width=0.48\textwidth]{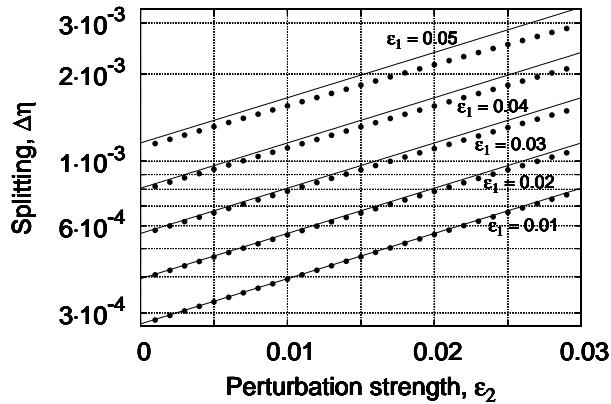}
\caption{Quasienergy splitting as a function of the strength of the second perturbation for different values of the first one. Perturbation period $T = 1$, kicks shift $dT = 0.4$. Lines - analytical formula, points - numerical results.}\label{fig:dE_multi}
\end{figure}
\begin{figure}[!h]
\centering
\includegraphics[angle = 270,width=0.48\textwidth]{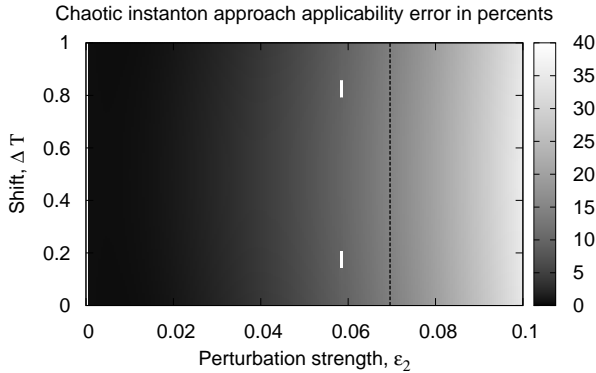}
\caption{Comparison error of the analytical formula~(\ref{an-fomula_multi}) with results of numerical calculations in percents. Perturbation period $T = 1$, perturbation strength $\epsilon_1 = 0.01$.}\label{fig:appl_multi}
\end{figure}

Using expressions for quasienergy splitting (\ref{eq:dn}), chaotic instanton action (\ref{eq:Sch}) and averaged Hamiltonian~(\ref{def:hameff_multi}) analytical formula for ground quasienergy splitting in the multiple kicked double well potential is expressed in terms of perturbation strength values
\begin{equation}\label{an-fomula_multi}
\Delta \eta(\epsilon) = \Delta E_0 \, e^{k \, (\epsilon_1 + \epsilon_2)},
\end{equation}
where coefficient $k$ is defined using expression (\ref{k}).

Obtained analytical formula (\ref{an-fomula_multi}) is checked in numerical calculations (see figures \ref{fig:dE_multi} and \ref{fig:appl_multi}). One period evolution in case considered has the following form:
\begin{equation}\label{evol.op_multi}
	U = e^{- \frac{i \hat{H}_0 (T - \Delta T)}{2}} e^{- i \epsilon_1 T \hat{x}^2} e^{- i \hat{H}_0 \Delta T} e^{- i \epsilon_2 T \hat{x}^2} e^{- \frac{i \hat{H}_0 (T - \Delta T)}{2}}.
\end{equation}

Figure \ref{fig:dE_multi} shows that formula  (\ref{an-fomula_multi}) can be used for the description of the ground quasienergy splitting dependence. The applicability parametric region of the formula is demonstrated on the figure \ref{fig:appl_multi}. Restriction (\ref{condition_multi}) is shown by dashed line.

\section{Polychromatic perturbation}\label{sec:aperiod}

Finally we will consider double kick system with different values of the perturbation period. System Hamiltonian is given by
\begin{align}
H & = \frac{p^2}{2 m} + a_0 x^4 - a_2 x^2 + \epsilon_1 T_1 x^2 \sum^{+ \infty}_{n = - \infty} \delta(t- n T_1) \nonumber \\ & + \epsilon_2 T_2 x^2 \sum^{+ \infty}_{n = - \infty} \delta(t - n T_2),\label{ham:aper}
\end{align}
where $T_1$ and $T_2$ are periods of two perturbations.

Averaged Hamiltonian can be calculated by averaging of the perturbed Hamiltonian over time for less common multiple $T_{lcm}$ of two periods $T_1$ and $T_2$
\begin{equation}\label{def:hameff_aper}
H_{av}  = \frac{1}{T_{lcm}} \int^{T_{lcm}}_0 H dt = \frac{p^2}{2 m} + a_0 x^4 - \left(a_2 - \epsilon_1 - \epsilon_2 \right) x^2.
\end{equation}

\begin{figure}[!b]
\centering
\includegraphics[angle = 270,width=0.48\textwidth]{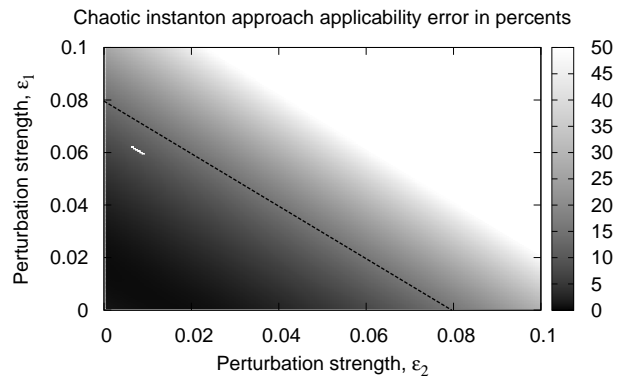}
\caption{Comparison error of the analytical formula~(\ref{an-fomula_multi}) with results of numerical calculations in percents. Perturbation periods $T_1 = 0.6$ and $T_2 = 1$.}\label{fig:appl_aperEps}
\end{figure}

\begin{figure}[!t]
\centering
\includegraphics[angle = 270,width=0.48\textwidth]{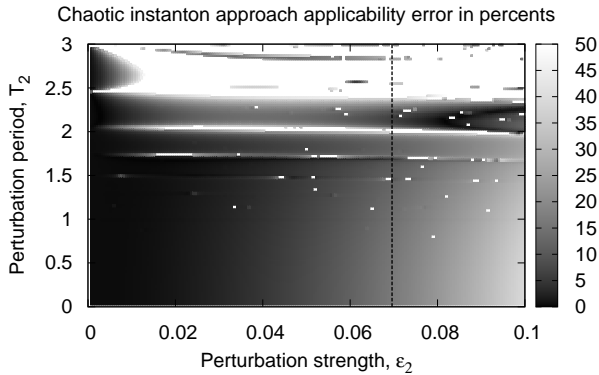}
\caption{Comparison error of the analytical formula~(\ref{an-fomula_multi}) with results of numerical calculations in percents. Parameters of the first perturbation $T_1 = 1$ and $\epsilon_1 = 0.01$.}\label{fig:appl_aper}
\end{figure}

Obtained averaged Hamiltonian is the same as for monochromatic multiple kicked system~(\ref{def:hameff_multi}). Thus analytical formula for ground quasienergy splitting will have the form (\ref{an-fomula_multi}).

To calculate quasienergy levels we construct evolution operator $U(T_{lcm})$ for period of time $T_{lcm}$ in analogy with expression (\ref{evol.op_multi}). Subsequent steps of numerical calculations are identical to the algorithm presented in the  section \ref{sec:num}.

Maps of the approach applicability for aperiodic kicked system are presented on  figures \ref{fig:appl_aperEps} and \ref{fig:appl_aper}. There is a good agreement between analytical and numerical results in a wide range of perturbation parameters. Chaotic instanton approach applicability restrictions $T \ll 2 \pi/{\omega_0}$ and~(\ref{condition_multi}) explain the obtained figures. The last restriction is shown by dashed line on figures.

\section{Conclusions}
\addcontentsline{toc}{section}{Conclusions}

Chaotic instanton approach allows to describe analytically the influence of the polychromatic perturbation on quantum properties of nonlinear systems. Double well system with single, multiple and polychromatic kicked perturbation is regarded in the paper to compare quantitative analytical predictions with the results of numerical calculations.

Chaotic instanton is the solution of the Euclidean equations of motion of the perturbed system. This configuration is responsible for the enhancement of tunneling far away from the exact (avoided) level crossings. Time-independent averaged system is used for regular approximation of the chaotic instanton solution in order to take into account its contribution to the ground quasienergy doublet splitting. Formula for the ground quasienergy levels splitting was derived averaging trajectory action in stochastic layer in the framework of chaotic instanton approach. This formula predicts exponential dependence of the ground doublet splitting on value of the perturbation strength. 

Numerical calculations for quasienergy levels dependence on value or values of single, multiple and polychromatic perturbation strength and simulations for tunneling dynamics are performed to check the validity of the obtained analytical formulas. Results of numerical calculations for the quasienergy spectrum  confirm the exponential dependence of the ground splitting on value of the perturbation strength for single perturbation or sum of values in multiple kicked case. They are in good agreement with the derived analytical formulas (\ref{an-fomula}) and (\ref{an-fomula_multi}). Simulations of the tunneling dynamics in the kicked double well system demonstrate exponential tunneling enhancement as well. Applicability of chaotic instanton approach was tested in a series of numerical calculations. Sufficiently wide range of perturbation parameters was found suitable for developed approach application.


\end{document}